\makeatletter
\declare@file@substitution{revtex4-1.cls}{revtex4-2.cls}
\makeatother

\documentclass[twocolumn]{aastex631}  

\usepackage{graphicx}
\usepackage{amsmath}
\usepackage{txfonts}
\usepackage{upgreek}

\usepackage{svg}

\begin{document}

   \title{Origin of Phobos and Deimos : Orbital evolution shortly after formation from a potential dislocation}

    \author{Ryan Dahoumane}
    \email{ryan.dahoumane@obspm.fr}
    \affiliation{LTE, Observatoire de Paris, Université PSL, Sorbonne Université, Université de Lille, LNE,CNRS \\
    61 Avenue de l’Observatoire, 75014 Paris, France}
    \author{Kévin Baillié}
     \affiliation{LTE, Observatoire de Paris, Université PSL, Sorbonne Université, Université de Lille, LNE,CNRS \\
     61 Avenue de l’Observatoire, 75014 Paris, France}
    \author{Valéry Lainey}
     \affiliation{LTE, Observatoire de Paris, Université PSL, Sorbonne Université, Université de Lille, LNE,CNRS \\
     61 Avenue de l’Observatoire, 75014 Paris, France}
    \accepted{}

   \begin{abstract}
   This paper deals with the formation and evolution of Mars' moons, Phobos and Deimos, assuming the dislocation of a larger progenitor as the origin of these moons. The study by \cite{Hyodo_2022} argue that under somewhat simplistic modeling, the post-dislocation  orbits of Phobos and Deimos inevitably collide within 10,000 years, leading to their mutual annihilation. These findings are based on $\mathcal{N}$-body simulations, accounting for Mars' $J_2$ and $J_4$ gravitational perturbations and mutual perturbations between the moons.
    In this paper, we challenge these findings by extending their work. We incorporate important perturbations such as solar perturbations, Mars' axial precession and nutation, and its deformation along three axes. We also extend some of the hypotheses made by \cite{Hyodo_2022} concerning the initial distribution of Phobos and Deimos after the dislocation. Our analysis reveals that including these additional perturbations as well as the possibility of having more than two fragments after the dislocation does not alter the ultimate fate of Phobos and Deimos. The moons still converge towards collision within comparable timescales, supporting \cite{Hyodo_2022} conclusions that the dislocation hypothesis under the dynamical scenario developed by \cite{Bagheri_2021} has, in the best conditions, about 10\% chance of surviving after the first 100,000 years following their formation.
    \end{abstract}
\section{Introduction} \label{sec:intro}
The origin of Phobos and Deimos is still a mystery yet to be unravelled. Two main scenarios have emerged from the studies carried out to date: direct capture of Phobos and Deimos and in-situ formation.
The arguments in favor of a direct separate capture of Phobos and Deimos usually rely on their shape and surface composition similarities with D or T-type asteroids (\citet{Rosenblatt2011TheOO}). However, considering the small size of Phobos and Deimos, a tidal capture seems very unlikely. Indeed, in such a scenario, both Phobos and Deimos orbits would get high eccentricities and inclinations just after their capture, as a capture on near circular and near-equatorial orbit is highly unlikely. Assuming that the change of orbits was mainly due to the tidal forces (as it is today), the lowering of their eccentricities and inclinations would require tides that are so strong that they do not seem compatible with what is thought to be the composition of Phobos and Deimos (\cite{Rosenblatt2011TheOO}). 

Another way to capture the satellites and lower their inclinations and eccentricities would be through dragging forces. Indeed, \cite{HUNTEN1979113} have shown that a rotating atmosphere of Mars would be able to capture satellites like Phobos and Deimos and make their orbit quickly evolve to their present positions. However, the main problem with this scenario is that it antedates the formations of Phobos and Deimos to ages when remnants of the rotating disc were still surrounding Mars. This scenario would leave no chance for Phobos to have survived until today considering its rapid descend due to tidal forces (\cite{Lainey2007}). This hypothesis however could be a valid explanation for Deimos origin, assuming the two satellites have different origins.

In-situ formation scenarios have also been proposed, for instance, by \cite{CRADDOCK2011} who defends the idea of a giant impact between Mars and a planetesimal. Indeed, this impact would have produced a debris disc from which Phobos and Deimos could have accreted. This scenario is compatible with a rubble-pile structure for Phobos and Deimos, and helps to explain the various Martian craters that would be due to fallout falls from the accretion disc. \cite{HesselBrock2017} proposed the hypothesis of re-accretion cycles for the formation of Phobos. Indeed, the proto-Phobos would cross the Roche limit, creating a ring of debris, from which 80\% of the mass would fall on Mars and the remaining 20\% would re-accrete to form a new generation of Phobos, and so on, diminishing slowly but surely the mass of the Mars innermost satellite. However, in a recent paper \cite{Madeira_2023} argued that the re-accretion cycles scenario would imply the existence of a ring close to the Roche limit, which should have been detected in all the previous martian observations.

Another \textit{in-situ} formation scenario proposed by \cite{Singer2007} describes Phobos and Deimos as the survivors of the fragmentation of a larger common progenitor satellite. Indeed, the capture of a more massive object is dynamically easier thanks to the greater effects of tides. The capture of this large satellite would have modified Mars' angular momentum, reducing its rotational speed to its current level. At the same time, as the progenitor approached the Roche limit, it would have fragmented, as a result of which the tidal forces would have pulled the more massive fragments towards Mars, leaving only Phobos and Deimos.

\section{The fragmentation model}
\subsection{Dislocation and migration}

\begin{figure*} [ht!]
  \centering
  \includegraphics[width=1.01
  \textwidth]{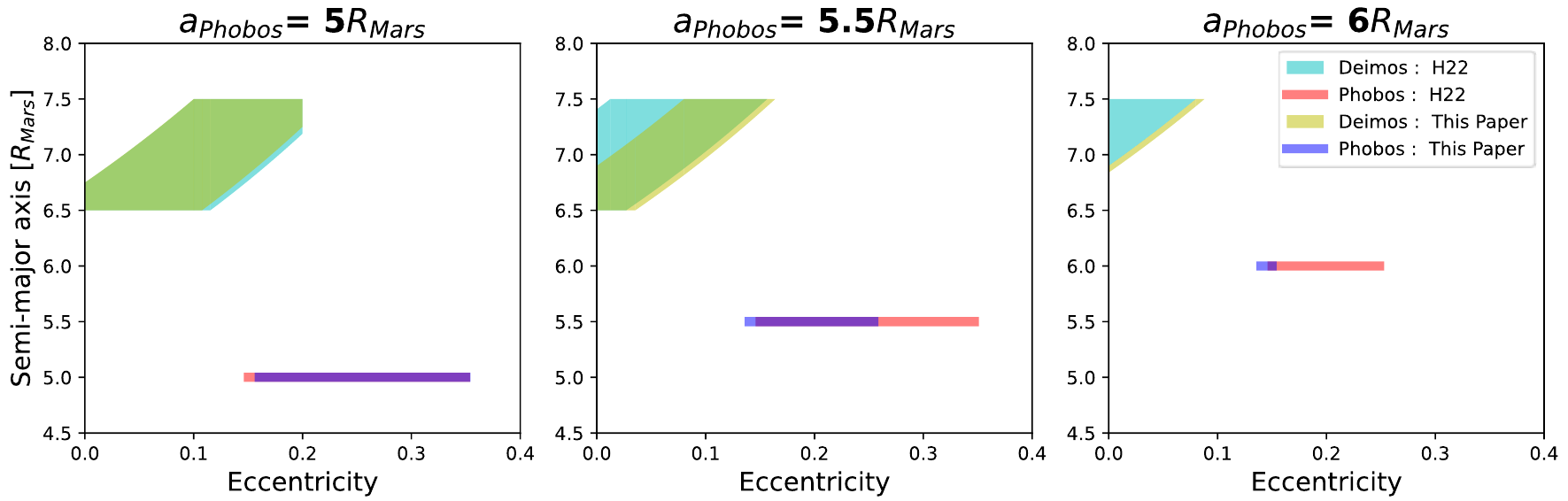}
  \caption{Distribution of the initial semi-major axes and eccentricities of the simulations. Each colored box represents the orbital elements distribution allowed by the constraints mentioned in Sec \ref{sec:new_eo}. 
  The "H22" refers to the one used by \cite{Hyodo_2022} in their simulations, while the "This Paper" shows those used in the present paper. Note that the green color is the superposition of yellow and light blue, while the purple is the overlap of red and dark blue.}
  \label{fig:eo_sat}
\end{figure*}

The scenario we mainly focus on in this paper is the one detailed in \cite{Bagheri_2021}, i.e. the formation of Phobos and Deimos by dislocation and subsequent migration of fragments of a common progenitor satellite.
By considering the Mars-Phobos and Mars-Deimos systems as separate systems, they have integrated backwards the semi-major axes, eccentricities and inclinations of the satellites from their present configuration. These integrations are based on a model of mutual tides between Mars and its satellites, under the assumption that the satellites are always locked in 1/1 spin-orbit resonance. In addition to these tidal forces, they added the physical libration of the satellites, which has the effect of increasing the dissipation due to the tides.
The authors estimated the quality factor $Q$ for Phobos based on viscosity estimates and granular friction studies of loose aggregates. The $k_2$ factor was calculated numerically, using a two-layer model consisting of a solid core and an outer porous layer, each as thick as half the satellite's radius. For Deimos, these two values were estimated by size-scaling the values of Phobos.

By integrating these equations over several billion years, \cite{Bagheri_2021} showed that Phobos may have been very close to Mars' synchronous orbit. The possible high eccentricity of Phobos at that time could even have made the satellite pass above the synchronous radius while having a semi-major axis lower to the synchronous value. This result can be explained by the tidal influence of Mars on Phobos, which has the systematic effect of reducing the satellite's orbit. If the orbits of Phobos and Deimos are taken far enough back in time, they cross near or beyond the synchronous orbit (more specifically between $5.9$ and $6.9 R_{Mars}$. These crossings would have occurred between 1 Gyr ago for the weakest cases ($k_{2_{\text{Pho}}}/Q_{\text{Pho}} = 1 \times 10^{-6}$ and $k_{2_{\text{Dei}}/Q_{\text{Dei}}} = 2 \times 10^{-4}$
) and 2.7 Gyr for the cases in which Phobos and Deimos are more consolidated ($k_{2_{\text{Pho}}}/Q_{\text{Pho}} = 6 \times 10^{-7}$ and $k_{2_{\text{Dei}}}/Q_{\text{Dei}} = 1.2 \times 10^{-4}$).

These observations therefore suggest a fragmentation scenario in which a progenitor satellite would generate Phobos and Deimos, which would have then migrated to their current positions thanks to the effects described above.

\subsection{Caveats}

\cite{Hyodo_2022} published a paper questioning the credibility of \cite{Bagheri_2021} conclusions. Using an $\mathcal{N}$-body simulation and the post-collision orbital elements of the satellites calculated by  \cite{Bagheri_2021}, \cite{Hyodo_2022} showed that within 10,000 years, the dynamical evolution of the system almost systematically results in a collision between Phobos and Deimos, leading to the total destruction of both objects.

They performed 1800 $\mathcal{N}$-body simulations of the Mars Phobos Deimos system using randomised orbital elements for Phobos and Deimos to estimate the probability of collision. The $\mathcal{N}$-body simulations are based on the integration of the equations of motion of the two satellites, which considered mutual gravitational perturbations and the effect of Mars flattening by mean of $J_2$ and $J_4$ zonal coefficients. In order to have initial conditions unlikely to generate collisions, the authors have considered only cases in which, after fragmentation, the orbits of both satellites have only one point of contact which corresponds to the apoapsis of Phobos and the periapsis of Deimos (i. e $a_{\text{Pho}} ( 1+e_{\text{Pho}}) = a_{\text{Dei}} (1-e_{\text{Dei}})$ and $|\omega_{Pho}-\omega_{Dei}| =\pi$ ). In spite of these considerations, a collision between both fragments happens in more than 94\%  of the simulations.
From impact simulations, the authors demonstrate that such collisions (100-300 $\text{m}\cdot \text{s}^{-1}$) would lead to annihilation of both satellites, creating a ring of debris. Based on these conclusions, it seems that the model proposed by \cite{Bagheri_2021} is unlikely.

\subsection{Limitations and further constraints}

\subsubsection{Post fragmentation orbital elements}\label{sec:new_eo}
Figure 1 from \cite{Bagheri_2021} as well as the conditions on the apsides described above allowed \cite{Hyodo_2022} to deduce the following constraints on the post-fragmentation orbital elements of Phobos and Deimos:
\begin{alignat*}{2}
a_{\text{Pho}} &\in [5 R_{\text{Mars}},6R_{\text{Mars}}] && \qquad
a_{\text{Dei}} \in [6.5 R_{\text{Mars}},7.5R_{\text{Mars}}] \\
e_{\text{Pho}} &\in [0.15,0.35] && \qquad
e_{\text{Dei}} \in [0,0.2] \\
i_{\text{Pho}} &= 0.021 \text{ rad} && \qquad
i_{\text{Dei}} = 0.015 \text{ rad}\\
\Omega_{\text{Dei}} &= \Omega_{\text{Pho}} && \qquad
|\omega_{\text{Dei}} - \omega_{\text{Pho}}| = \pi \text{  rad} \\
a_{\text{Pho}}&(1+e_\text{Pho}) = a_{\text{Dei}}(1-e_\text{Dei})&& \qquad
\end{alignat*}
where $a$,$e$,$i$,$\omega$, $\Omega$ are respectively the semi-major axes, eccentricities, inclinations, arguments of the periapsis and longitudes of the ascending node. This orbital elements distribution can be seen in Fig. \ref{fig:eo_sat}

We can make several comments on the choice of these constraints:
\begin{itemize}
    \item[-] Figure 1 by \cite{Bagheri_2021}, showing the evolution of the eccentricity of Phobos and Deimos, actually indicates a minimum post-fragmentation eccentricity of Phobos of slightly less than 0.15. For their simulations, \cite{Hyodo_2022} should therefore probably have distributed the eccentricities of Phobos between 0.14 and 0.35 in order to be conservative.
    \item[-] It is explicitly stated in \cite{Bagheri_2021} that the orbits intersect at a distance of 5.9 to 6.9 Mars radii. This constraint does not seem to have been taken into account by \cite{Hyodo_2022} in their distribution.
\end{itemize}

This second point significantly changes the distribution of the initial orbital elements of the two moons, in particular by greatly reducing the mean eccentricity of Deimos. Note that circular orbits increase the chances of collisions, since zero eccentricity virtually eliminates the relative precession of the apsids, and only the precession of the nodes can separate the orbits. From these new orbital distributions presented in Fig. \ref{fig:eo_sat}, we can expect a drop in collision probability, particularly for high values of Phobos' semi-major axis, when this new constraint drastically reduces the number of circular orbits.

Furthermore, in order to minimise the chances of collision, \citet{Hyodo_2022} had assumed that the apoapsis of Phobos is equal to the periapsis of Deimos. This is a good idea, however we will consider a situation that minimises further the chances of collisions by slightly increasing the distance between satellites in the initial conditions. Instead of making the centre of the satellites coincide using the condition $a_{\text{Pho}}(1+e_\text{Pho}) = a_{\text{Dei}}(1-e_\text{Dei})$ , we will consider the case where the two moons are barely in contact after fragmentation: $a_{Pho}(1+e_{Pho}) + R_{Pho} = a_{Dei}(1-e_{Dei}) - R_{Dei}$, with $R$ being the radius of each moon.
In this way, we expect the number of collisions at the start of the simulation to be greatly reduced, as any small perturbation would prevent them. This new orbital elements distribution is shown in Fig. \ref{fig:eo_sat} and is based on the following constraints :
\begin{align*}
a_{\text{Pho}} &\in [5 R_{\text{Mars}},6R_{\text{Mars}}] && \qquad
a_{\text{Dei}} \in [6.5 R_{\text{Mars}},7.5R_{\text{Mars}}] \\
e_{\text{Pho}} &\in [0.14,0.35] && \qquad
e_{\text{Dei}} \in [0,0.2] \\
i_{\text{Pho}} &= 0.021 \text{ rad} && \qquad
i_{\text{Dei}} = 0.015 \text{ rad} \\
\Omega_{\text{Dei}} &= \Omega_{\text{Pho}} && \qquad
|\omega_{\text{Dei}} - \omega_{\text{Pho}}| = \pi \text{ rad}
\end{align*}
\[5.9 R_{\text{Mars}} \leq a_{\text{Pho}}(1+e_\text{Pho}) + R_{\text{Pho}} = a_{\text{Dei}}(1-e_\text{Dei}) - R_{\text{Dei}} \leq 6.9 R_{\text{Mars}}
\]

\subsubsection{Limited physical effects} 
In addition, we propose to increase the complexity of the physical model used by \cite{Hyodo_2022} in their paper to test the robustness of their results.  The authors justified the simplicity of their model by explaining that the most important perturbation apart from the flattening of Mars is evection, which was said to have no impact on the result. We propose to verify this claim. We extended \cite{Hyodo_2022} model by adding the gravitational perturbations due to the Sun, the dynamical triaxiality of Mars (coefficients $c_{22}$ and $s_{22}$), and the precession and nutation of Mars.

Solar perturbations are expected to influence the nodal precession of satellite orbits, particularly when they are orbiting far from Mars. This effect is expected to intensify when increasing Martian obliquity. Additionally, the triaxial shape of Mars may induce resonant interactions with the satellites when they approach spin-orbit resonance. Furthermore, the precession of Mars’ rotational axis could induce periodic variations in the satellites’ orbital inclinations. However, this influence is likely to be minor, as the precession of the satellites’ ascending nodes occurs at a faster rate than the planetary precession (see \cite{Goldreich1965InclinationOS} and \cite{gurfil_2006}). 

For the sake of completeness, we also added the $J_3$ flattening term for Mars. This term was neglected by \cite{Hyodo_2022} because the low inclinations of Phobos and Deimos make the $J_3$ perturbations negligible.  

Additionnaly, we included tidal effects in our modeling. Within such short timescales (tens of thousands of years), the orbital change due to tides within Mars as well as within Phobos and Deimos are expected to be negligible. However, in order to be as exhaustive as possible, we have conducted additional simulations where we considered extreme values for the $k_2/Q$ of Mars, Phobos and Deimos significantly higher than present values.
We neglect the relativistic effect as the relativistic shift of Phobos periapsis is about $3 \cdot 10^{-7}$ $\text{rad}\cdot \text{year}^{-1}$ based on \cite{arrighi_2015_discrete} ($\Delta\omega=\frac{24 \pi^3 a^2}{T^2 c^2\left(1-e^2\right)}$ with $\Delta\omega$, $c$ and $T$ respectively the shift in periapsis, the speed of light and the Newtonian orbital period). As a comparison, based on the Eq. (14) from \cite{Hyodo_2022}, the precession of Phobos periapsis due to Mars oblateness is about $0.7$ $\text{rad}\cdot \text{year}^{-1}$, five to six orders of magnitude greater.

\section{Methodology \label{Methodo}}
\subsection{Dynamical modeling} 
\subsubsection{Equations of motion}

In a planetocentric frame, the acceleration undergone by each moon considered as point masses can be written as (\cite{Lainey2004}) :

\begin{equation}
\begin{aligned}
\ddot{\mathbf{r}}_i= & -\frac{G\left(m_0+m_i\right) \mathbf{r}_i}{r_i^3} \\
& +\sum_{j=1, j \neq i}^{\mathcal{N}} G m_j\left(\frac{\mathbf{r}_j-\mathbf{r}_i}{r_{i j}^3}-\frac{\mathbf{r}_j}{r_j^3}\right)+G\left(m_0+m_i\right) \nabla_i V_{\bar{\imath} \hat{0}}
\label{eqn:Newton}
\end{aligned}
\end{equation}

The subscripts $i$ and $j$ represent any of the $\mathcal{N}$ bodies in the problem, with the exception of the central body, which is symbolised by the subscript $0$. $G$ is the universal gravitational constant, $\mathbf{r_i}$ represents the position vector of body $i$ in the reference frame described above, $r_i$ represents the norm of this vector, and $r_{ij}$ represents the distance between body $i$ and body $j$. $\nabla$ designates the gradient operator and $V_{\bar{\imath} \hat{0}}$ represents the gravitational term due to the shape of the central body exerted on the $i$ body. As in \cite{Lainey2004} , we divided this last term in two parts: $V_{\bar{\imath} \hat{0}} = V_{\bar{i} \hat{0}}^{(1)}+ V_{\bar{i} \hat{0}}^{(2)}$ where $V_{\bar{i} \hat{0}}^{(1)}$ is the term due to the flattening of the central body and $V_{\bar{i} \hat{0}}^{(2)}$ is the term due to the sectorial terms.

The derivative of the term due to the flattening $J_2$, $J_3$ and $J_4$ can be written as follows (Eq. (1.4) from \cite{Lainey2004}):

\begin{equation}
\begin{aligned}
& \nabla_i V_{\bar{i} \hat{0}}^{(1)}= \\
& -\sum_{n=2}^{\infty} \frac{(R)^n J_n}{r_i^{n+2}}\left[\left(\mathbf{k}-\frac{\mathbf{r}_{\mathbf{i}}}{r_i} \sin \phi_i\right) P_n^{\prime}\left(\sin \phi_i\right)-(n+1) \frac{\mathbf{r}_{\mathbf{i}}}{r_i} P_n\left(\sin \phi_i\right)\right]
\end{aligned}
\end{equation}

Expanding the expression to order 4, we find :

\begin{equation}
\begin{aligned}
\nabla_i V_{\bar{i} \hat{0}}^{(1)}=\frac{\left(R\right)^2 J_2}{r_i^{4}}\left[\frac{ \mathbf{r_i}}{r_i}P'_3\left(\sin \phi_i\right)  - \mathbf{k} P'_2\left(\sin \phi_i\right) \right] \\
+ \frac{\left(R\right)^3 J_3}{r_i^{5}}\left[\frac{ \mathbf{r_i}}{r_i}P'_4\left(\sin \phi_i\right)  - \mathbf{k} P'_3\left(\sin \phi_i\right) \right]\\
+ \frac{\left(R\right)^4 J_4}{r_i^{6}}\left[\frac{ \mathbf{r_i}}{r_i}P'_5\left(\sin \phi_i\right)  - \mathbf{k} P'_4\left(\sin \phi_i\right) \right]\\
\end{aligned}
\label{eqn:flattening}
\end{equation}
With $\mathbf{k}$ the rotation vector of the central body, $R$ the equatorial radius of the central body, $\phi_i$ the latitude of the body $i$ with respect to the equator of the central body, $P_n$ the Legendre polynomial of degree $n$, and $P'_n$ the derivative of the Legendre polynomial of order $n$ with respect to $\sin{\phi}$. The sine of the latitude can easily be found as follows:
\begin{equation}
    \sin(\phi_i) =\frac{\mathbf{r_i \cdot k}}{r_i}
\end{equation}

\begin{figure} [!t]
  \centering
  \includegraphics[width=0.5
  \textwidth]{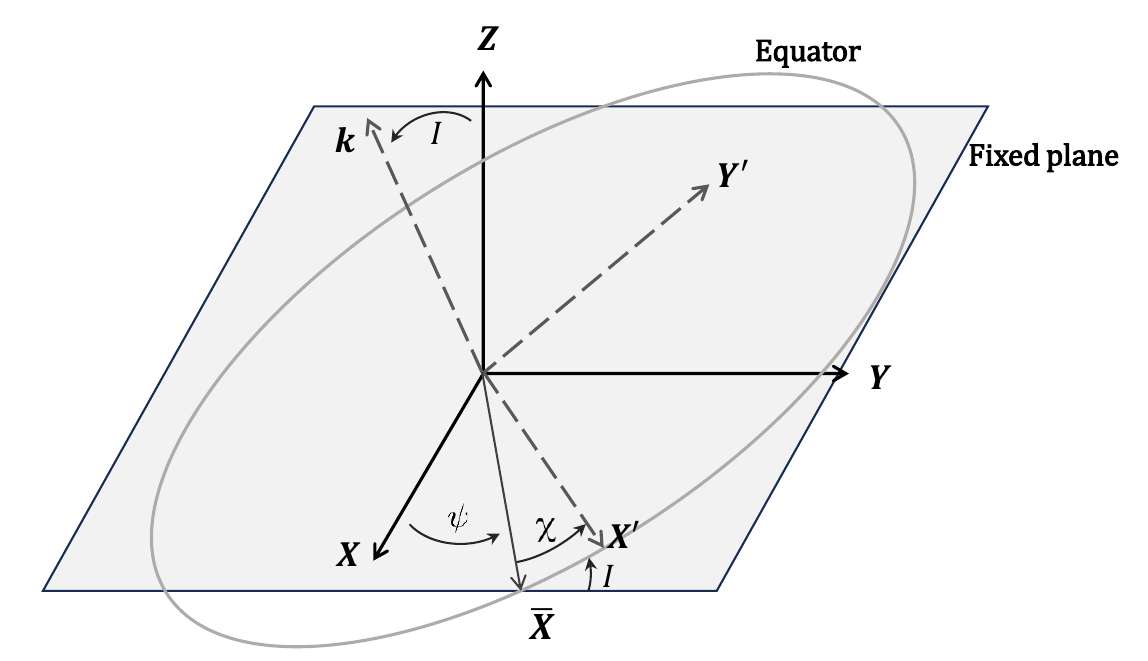}
  \caption{Representation of the fixed and moving frames. The first frame which is fixed is defined by arbitrary axes $(\mathbf{X},\mathbf{Y},\mathbf{Z})$. After three rotations with the angles $(I,\psi, \chi)$, this fixed frame can be turned into the rotating frame $(\mathbf{X'},\mathbf{Y'},\mathbf{k})$ where $\mathbf{X'}$ is toward the first meridian and $\mathbf{k}$ is the perpendicular to the equator. We also represented $\mathbf{\Bar{X}}$ which is directed toward the ascending equatorial node.
  }
  \label{fig:reperes}
\end{figure}

As mentioned previously, we decided to include the term related to the $J_3$ coefficient in the calculation of the potential for the sake of completeness.
However, even without this term, we have slightly different expression from that given by \cite{Hyodo_2022} for three reasons.
\begin{itemize}
    \item[] First, the expression for the potential due to the flattening of the central body is different (Eq. \ref{eqn:flattening}). This is because the authors assumed a fixed rotation axis and thus defined a reference frame $xyz$ with $xy$ the Martian equatorial plane and $z$ the rotation axis of the central body. Given that we are going to consider the precession and nutation of the central body's rotation axis, we are not in the same type of frame and therefore have a different, more general expression. Indeed, by developing Eq. \ref{eqn:flattening}, considering $\mathbf{k}=(0,0,1)^T$ and neglecting the term due to $J_3$, we end up with the same expression of the potential due to the flattening of the central body as that given by \cite{Hyodo_2022}. 

    \item[] Secondly, they have neglected the mass of the body $i$ when summing it up with the mass of Mars in the first term of the Eq. \ref{eqn:Newton}. This could be done, as Phobos and Deimos masses are negligible when compared to Mars. However, as we include the Sun in our modeling, $m_i$ can not be neglected in this paper.

    \item[] Finally, it also appears that the authors omitted the indirect terms in their calculation of the force. Fortunately, this omission does not seem to affect the results since, as can be seen from figures \ref{fig:cc_full} and \ref{fig:vit_coll}, the results of their simulations can be found from our formulation of the force, which includes the indirect terms.

\end{itemize}
Let us now express the derivative of the term depending on $c_{22}$ and $s_{22}$, due to the disparities in longitude of the Mars mass distribution (sectorial terms). This can be written as (\cite{Lainey2004}) :\\

\begin{equation}
\begin{aligned}
\nabla_i V_{\bar{i} \hat{0}}^{(2)}= & \frac{(R)^2}{r_i^4}\left[\frac{\mathbf{r}_{\mathbf{i}}}{r_i}\left(15 \sin ^2 \phi_i-9\right)-6 \mathbf{k} \sin \phi_i\right] \\
& \times\left[c_{22} \cos 2 \lambda_i+s_{22} \sin 2 \lambda_i\right] \\
+ & \frac{(R)^2}{r_i^3} 6\left(1-\sin ^2 \phi_i\right) \frac{x_i^{\prime} \nabla y_i^{\prime}-y_i^{\prime} \nabla x_i^{\prime}}{x_i^{\prime 2}+y_i^{\prime 2}} \\
& \times\left[-c_{22} \sin 2 \lambda_i+s_{22} \cos 2 \lambda_i\right]
\end{aligned}
\end{equation}

Where $\lambda_i$ is the longitude of the position of the body $i$ and $(x_i', y_i', z_i')$ the coordinates of the body $i$ in the rotating frame centred on the central body $(P_0, \mathbf{X'}, \mathbf{Y'}, \mathbf{k})$, with $x_i'$ the equatorial coordinate pointing towards the first meridian, $z_i'$ the coordinate pointing along the axis of rotation of the central body and $y_i'$ completing the direct trihedron (see Fig.\ref{fig:reperes}).
The transition matrix $M$ between the coordinates of this rotating frame and the fixed frame $(P_0, \mathbf{X}, \mathbf{Y}, \mathbf{Z})$ such that $(x_i', y_i', z_i')^T = M(x_i, y_i, z_i)^T$ is the following (\cite{Lainey_these}, \cite{Lainey2004}) :

\begin{equation}
\resizebox{\linewidth}{!}{\arraycolsep=2pt%
$M=\left(
\begin{array}{ccc}
\cos \chi \cos \psi-\sin \chi \sin \psi \cos I & \cos \chi \sin \psi+\sin \chi \cos I \cos \psi & \sin \chi \sin I \\
-\sin \chi \cos \psi-\cos \chi \sin \psi \cos I & -\sin \chi \sin \psi+\cos \chi \cos \psi \cos I & \cos \chi \sin I \\
\sin I \sin \psi & -\cos \psi \sin I & \cos I
\end{array}\right)
$}
\end{equation}
with $I$, $\psi$ and $\chi$ being respectively the inclination of the rotation axis (obliquity), the longitude of the ascending node and the angle between the ascending node and the prime meridian .

\subsubsection{Computing the precession-nutation of Mars}
The precession and nutation motions of a body orbiting the Sun are given by \cite{gurfil_2006}. It is possible to describe the evolution of Mars' axis of rotation as follows:
\begin{equation}
    \frac{d \mathbf{k}}{d t}=\alpha(\mathbf{n} \cdot \mathbf{k})(\mathbf{k} \times \mathbf{n})
\end{equation}
where $\mathbf{k}$ denotes the Martian spin axis, $\mathbf{n}$ the normal to the orbital plane and $\alpha$ a parameter proportional to the oblateness factor $J_2$. The $\mathbf{n}$ vector can be written as 
\begin{equation}
    \mathbf{n}= \left(q, -p, \sqrt{ 1-p^2 -q^2}\right) ^T
\end{equation}

with 

\begin{equation}
q=\sin I_{\text {orb }} \sin \Omega_{\text {orb }}  \quad p=\sin I_{\text {orb }} \cos \Omega_{\text {orb }}    
\end{equation}
where $I_{\text {orb }}$ is the inclination of the Martian orbit with respect to a fixed plane and $\Omega_{\text {orb }}$ the ascending node with respect to this same plane.
It is trivial to calculate these values directly from the relative positions and velocities of Mars and the Sun, however, for the sake of precision, similarly to \cite{gurfil_2006}, we use \cite{Ward1979}'s model which describes the evolution of these angles as a function of time in the fixed reference frame B1950. We therefore use Eq. 20 and 21 from \cite{gurfil_2006} to calculate the coefficients $q$ and $p$.

\subsubsection{Tides in Mars}
To take into consideration the effect of martian tides on Phobos and Deimos, we have used the expression given in Equation (3) from \cite{Lainey2007} which gives the force acting on a moon induced by the bulged created by this moon on the planet. Considering that Mars is the center of our referential, we can write

\begin{equation}
\label{eqn:tide_planete}
\ddot{\mathbf{r}}^{\mathrm{T_0}}_{i}=-\frac{3 k_2 G m_i\left(R\right)^5}{r_i^8}\left(\mathbf{r_i}+\Delta t\left[\frac{2 \mathbf{r_i}(\mathbf{r_i} \cdot \mathbf{v_i})}{r_i^2}+(\mathbf{r_i} \times \boldsymbol{\uprho}+\mathbf{v_i  })\right]\right)
\end{equation}

with  $k_2$, $\Delta t$ respectively the planet's Love number, the time shift between the excitation of the planet and its viscoelastic response. $\boldsymbol{\uprho}$ represents the rotation vector of the planet such that

\begin{equation}
    \boldsymbol{\uprho} = \mathbf{k}\dot\chi 
\end{equation}

Moreover, Equations (1) and (2) from \cite{Lainey2007} give us 

\begin{equation}
    \Delta t = \frac{T^{rot}_0 T_i \arcsin{(1/Q)}}{4\pi \lvert{T^{rot}_0-T_i}\rvert}
\end{equation}

with $T^{rot}_0$, $T_i$ and $Q$ the planet's rotation period, the satellite's revolution period around the planet and the planet's tidal quality factor.

In this work, we will assume $k_2$ and $Q$ as constants, as the point is to check if extreme values can alter the results rather than having an accurate representation of the phenomenon. Moreover, because of the small size of the satellites compared to the planet, we neglect the influence of Phobos and Deimos on the Mars rotation speed $\dot\chi$ .

\subsubsection{Tides in Phobos and Deimos}
Finally, the last effect we added in the modeling of our problem is the force exerted by the moons on the planet due to the tidal bulges of the moons. In an analog way to Eq. \ref{eqn:tide_planete}, the acceleration of the body $i$ due to this force in a planetocentric referential can be written 

\begin{equation}
\label{eqn:tide_moon}
\ddot{\mathbf{r}}^{\mathrm{T_i}}_{i}=-\frac{3 k_{2i} G m_0\left(R_i\right)^5}{r_i^8}\left(\mathbf{r}_i+\Delta t\left[\frac{2 \mathbf{r}_i(\mathbf{r}_i \cdot \mathbf{v}_i)}{r_i^2}+(\mathbf{r}_i \times \boldsymbol{\uprho}_i+\mathbf{v  }_i)\right]\right)
\end{equation}
and 

\begin{equation}
    \Delta t_i = \frac{T^{rot}_i T_i \arcsin{(1/Q_i)}}{4\pi \lvert{T^{rot}_i-T_i}\rvert}
\end{equation}

with $k_{2i}$, $Q_i$,$R_i$, $\boldsymbol{\uprho}_i$,$T^{rot}_i$, $\Delta t_i $ respectively the Love number, tidal quality factor, equatorial radius, rotation vector and time lag of the body $i$ (satellite). We still consider $k_{2i}$ and $Q_i$ as constant for the same reasons we did with $k_{2}$ and $Q$. As we want to maximize the effect of the tides in the satellite, we will consider that, at any given time, the rotation of the satellites are in the same plan as their revolution around the planet.

Unlike the planet, due to their small size, the variation of the rotation speed of the satellites can not be neglected when computing the acceleration due to their tidal deformation. In other words, we can not consider the rotation speed $\rho_i$ of the satellite as constant. For this reason, we will be simultaneously integrating the position rotation of the satellites. To do so, we will be using the Equation (7) from \cite{Aleshkina2009}

\begin{equation}
-\frac{\mathrm{d} \rho_i}{\mathrm{~d} t}=K_{t,i}\left(\frac{1}{r_i}\right)^6+K_{l,i}\left(\frac{a_i}{r_i}\right)^3 \sin 2(\theta_i-f_i)
\end{equation}
with $a_i$ the semi-major axis of the satellite, $f_i$ its true anomaly and $\theta_i$ the angle between the axis of the moment of inertia $A$ of the satellite and its eccentricity vector ($\rho_i = \frac{\mathrm{d} \theta_i}{\mathrm{~d} t}$). The first term of the expression is due to the tidal bulge, while the second term is related to the inertia of the satellite. If we considered satellites as point masses, the second term would not appear. However, as Phobos and Deimos are not spherical, it must be taken into account. The expressions for the tidal coefficient $K_{t,i}$ and the inertial coefficient $K_{l,i}$ given in \cite{Aleshkina2009} are :

\begin{equation}
\label{coeff_maree}
K_{t,i}=\frac{15}{4} \frac{G m_0^2 R_i^3}{m_i Q_i} k_{2i}
\end{equation}
\begin{equation}
K_{l,i}=\frac{3}{2} \frac{B-A}{C} n_i^2 
\end{equation}
with $A$, $B$ and $C$ the main moments of inertia of the satellite and $n_i$ its mean motion.
As in \cite{Aleshkina2009}, we will assume that $C/m_iR_i^2 \approx0.4$. Moreover, the $c_{22}$ term of the satellite can be written (\cite{Yoder1995}) $c_{22} = \frac{1}{4} \frac{B-A}{m_iR_i^2}$. From this, the inertial coefficient can be rewritten as 

\begin{equation}
\label{coeff_inertial}
K_{l,i}\approx15 c_{22} n_i^2 
\end{equation}

\subsection{Numerical integration : Hybrid Radau integrator}

We now aim at solving numerically the equations of motion described above. In order to combine the speed of Everhart's original RA15 code \citep{RA15_everhart} with the improved accuracy provided by Rein and Spiegel's IAS \citep{REINIAS15}, we developed a hybrid integrator that can switch methods between RA15 and IAS at each time step depending on the distance between the different bodies within an $\mathcal{N}$-body simulation. This means that when the bodies are far apart, the classical Radau is used for fast integration. However, when two bodies are sufficiently close, the integrator used is the IAS, which guarantees a higher accuracy for close encounters and ensures that no collisions are missed.

To do so, after each time step, a check is made on the distance between each pre-selected bodies. If the distance between these objects is less than a threshold $\epsilon$, the integration of the next time step is made by the IAS, otherwise the Radau is used. We have set the threshold to activate the IAS when the lighter body is susceptible to enter the Hill sphere of the heavier. 

Let the $i$ subscript symbolise the heavier body and the $j$ subscript symbolise the lighter one. Let $\mathbf{V_{j \rightarrow i}}$ be the velocity vector of $j$ relative to $i$ and $r_{hill}(i)$ the radius of the Hill sphere of the body $i$. Assuming both bodies are spherical, let $R_i$ and $R_j$ be the radii of each body and $d$ the distance between them. Then, considering that the velocity magnitude is constant during the timestep, we can set 

\begin{equation}
        \epsilon = r_{Hill}(i)+(R_i +R_j) + \lVert {\mathbf{V_{j \rightarrow i}}} \rVert *(t_{n+1} -t_n)
\end{equation}

\subsection{Detecting collisions in the simulation}

Collision detection plays a major role in the simulations. The most straight-forward way to detect a collision between two bodies in an $\mathcal{N}$-body problem consists in checking the distance $d$ between them at each timestep : 
\begin{equation}\label{eqn:coll_cond}
d \leq R_i +R_j \implies \mathit{collision}
\end{equation}
This basic method presents however a flaw : it would miss a collision that occurs between two successive time steps. Nevertheless, our testing indicates that the IAS part of the integrator we are using is very efficient in reducing the timestep when during close approach, so we will not need any additional collision detection condition. In addition we want to avoid false positives as much as possible, as our simulations aim to set a higher bound on the survival rate.

\begin{table}[htb!]
    \centering
    \caption{Physical values used in this paper.}
    \begin{tabular}{llll}
    \hline \hline
        Variable & Value & Unit &Source \\ \hline
        $J_2$ & $1.96 \times 10^{-3}$ & $\emptyset$ &  \\ 
        $J_3$ & $3.15\times 10^{-5}$ & $\emptyset$ & \\ 
        $J_4$ & $-1.54\times 10^{-5}$ &$\emptyset$ & (1) \\ 
        $c_{22}$ & $-5.46322410\times 10^{-5}$ & $\emptyset$ &\\ 
        $s_{22}$ & $3.15871652\times 10^{-5}$ &  $\emptyset$ & \\ \hline
        $\alpha$ & $1.0878850102669404\times 10^{-7}$ & $\text{rad}\cdot\text{day}^{-1}$&  \\ 
        I & $25.25797549$ & deg & (2)\\ 
        $\psi$ & $332.6841708$ & deg &  \\ \hline
        $\chi$ & $[0,2\pi] + 350.89198d$ & deg & (3) \\ \hline
    \end{tabular}
    \tablecomments{$I$, $\psi$ and $\chi$ angles are given in B1950 \\(1) \citet{Genova_GMM3}; (2) \citet{gurfil_2006}; \\(3) \citet{Davies1983}}
    \label{tab:Valeurs_simu}
\end{table}
    \begin{figure*} [htb!]
      \centering
      \includegraphics[width=1
      \textwidth]{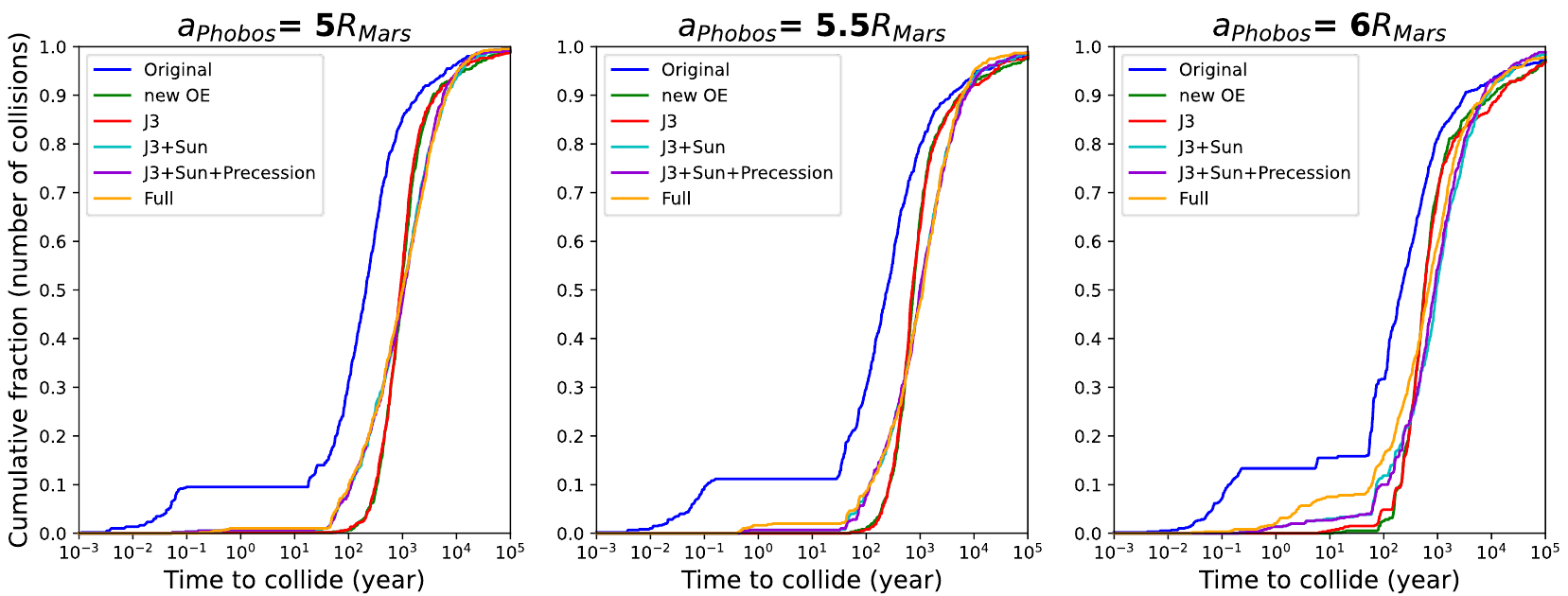}
      \caption{Cumulative fraction of collisions as a function of simulation time. We made different simulations considering different configurations. For each of this configurations, we made 600 simulations for each value of $a_{Phobos}$. The "Original" curves are a pure reproduction of  \cite{Hyodo_2022} results, "New OE" curves have the same physical effects as those considered by Hyodo et al., but with the new initial orbital elements repartition detailed in this paper. Then in the "J3" curves, we also added the oblateness term $J_3$, after which we also added the sun gravitational perturbations in the "J3 + sun" curves. As mentioned in its name, we then added the precession in the "J3+Sun+Precession" curves, and finally, we added the martian sectoral terms $C_{22}$ and $S_{22}$ in the "Full" simulation. 
      } 
      \label{fig:cc_full}
    \end{figure*}
\section{Results \label{results}}
\subsection{Changing the initial distribution and extending the simulation time }

    We first want to see how our new orbital elements distribution (described in Section \ref{sec:new_eo}) affect the collision probability while using the same physical process as \cite{Hyodo_2022}. By comparing the blue and the green curves from Fig. \ref{fig:cc_full}, we notice that, despite notable differences at the beginning of the simulation, we obtain a result very similar to that of \cite{Hyodo_2022} after 100,000 years of evolution: more than 94\% of the simulations have undergone a collision. However, there is a very sharp drop in the number of collisions in the short term, which can be explained by the fact that we have shifted the positions of the orbits of Phobos and Deimos so that they are separated by the sum of their physical radii. This new separation prevents the two bodies from colliding in their first revolution, but the mutual perturbations of Phobos and Deimos have the secular effect of bringing their orbits closer together, leading to a similar result as the one obtained by \cite{Hyodo_2022}.
    
    As for the effect induced by the new criterion (fragmentation distance between 5.9 and 6.9 Mars radii), which was omitted by Hyodo et al, it has a very weak effect on the evolution of the number of collisions, but has a somewhat stronger effect on the cases where $a_{Pho} = 6 R_{Mars}$. In this case, the \cite{Hyodo_2022} conditions tend to generate lower Deimos eccentricities than those of the new configuration. Calculating the median Deimos eccentricity for the 6 Mars radii case typically gives a value of around 0.02 for the \cite{Hyodo_2022} distribution, compared with a value of around 0.04 for the new distribution. It is possible that this reason is partly responsible for the reduction in the amplitude of the first collision peak in the new distribution of orbital elements in the case where $a_{Pho} = 6 R_{Mars}$.
    
    Despite this, it seems that a distribution even more unfavorable to collisions than the previous one is not enough to significantly reduce the probability of Phobos and Deimos colliding as the blue and green curves from Fig.\ref{fig:cc_full} are very close to each other after $10^5$ years.
\subsection{Adding physical effects}
    We now investigate the impact of each of the physical effects added to the simulation (results available in Fig. \ref{fig:cc_full}.)\\    
    As expected, given the very low inclinations of Phobos and Deimos, the effect of $J_3$ is negligible (green and red curves are almost undistinguishable). The perturbation potential due to the $J_3$ term on a $j$ body can be written
    $U_{J_3} = -\frac{\left(R\right)^3 J_3}{r_j^{5}}\left[ 5 sin^3(\phi_j)-3sin(\phi_j)\right]$. Knowing that \mbox{$sin(\phi_j) = sin(i_j)sin(\nu_j+\omega_j)$ with $i_j$, $\nu_j$ and $\omega_j$} respectively the inclination, the true anomaly and the argument of periapsis of the orbit of $j$. Each term of the potential due to $J_3$ therefore depends on the sine of the inclination, so $i \approx 0 \implies \nabla U_{J_3} \approx 0$.

    The addition of the Sun in the simulations (light blue curve) seems to create perturbations which tends to increase the number of collisions on the short term (approximately equal to 1 year). The effect of the solar perturbations increases with the semi major axes of Phobos and Deimos.
    
    The precession and nutation of Mars, on the other hand, have no significant effect on collisions. This can be explained by a very slow precession (about $2 \times 10^{-3} \text{deg}\cdot \text{year}^{-1}$, or 203 deg in 100,000 years, less than one complete revolution) and a fairly weak nutation on this timescale (oscillation between 23 and 27 degrees).
    
    Finally, taking into account the potential due to the sectorial terms $c_{22}$ and $s_{22}$ has a negligible effect, except for $a_{phobos} = 6 R_{Mars}$ (yellow curve). This is the configuration in which Phobos is very close to the synchronous orbit, which most likely induces a resonance that strongly disturbs its orbit.
    So it seems that the addition of all these physical effects is not sufficient to alter the chances of collisions between Phobos and Deimos after 100,000 years.
    
\begin{figure} [t!]
      \centering
      \includegraphics[width=0.49
      \textwidth]{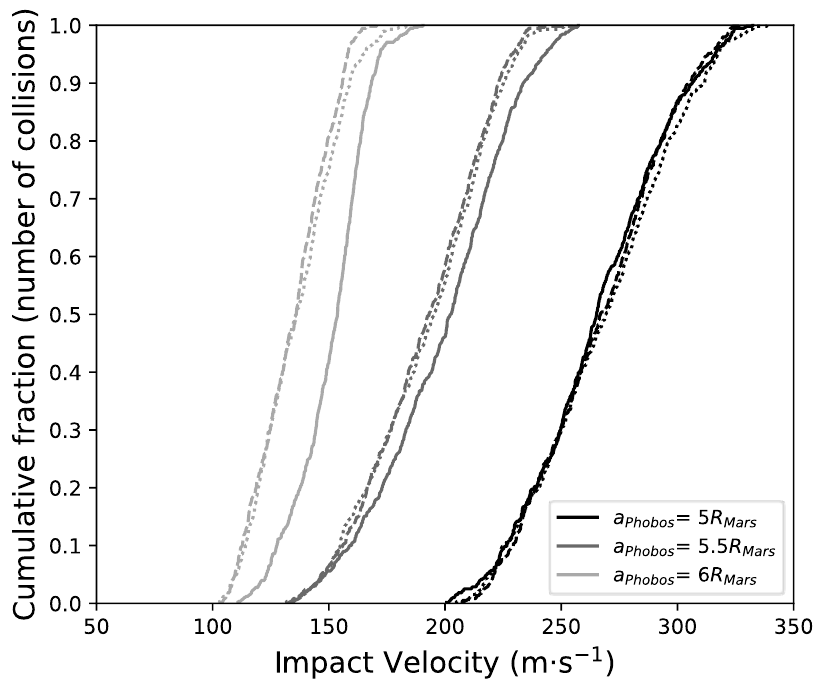}
      \caption{Cumulative fraction of collisions as a function of impact velocity. We did the same plot as in \cite{Hyodo_2022} to compare their results with our new simulations. These results come from the same simulations as those presented in Fig. \ref{fig:cc_full}. 
      The plain, hashed and dotted lines represent represent respectively the reproduction of \cite{Hyodo_2022} results, the results from a physical model similar to \cite{Hyodo_2022} but with new orbital elements (cf Fig. \ref{fig:eo_sat}), and the simulations with our new physical modeling and orbital elements distribution. The colors light, middle and dark gray represent respectively the simulations in which we considered the initial semi-major axis of Phobos as equal to 6, 5.5 and 5 Mars radii. 
      }
      \label{fig:vit_coll}
    \end{figure}

    Moreover, as visible in Fig. \ref{fig:vit_coll}, the new configuration does not significantly alter the impact velocity, thus, as in \cite{Hyodo_2022}, the collisions are destructive

\subsection{The chaotic obliquity of Mars}
     \begin{table}[!t]
        \centering
        \caption{ Mars orbital elements used for the simulations with varying obliquity.}
        \begin{tabular}{lll}
        \hline \hline
        Orbital Element & Value & Unit\\
        \hline
            Semi-major axis & 1.5237 & au\\ 
            Eccentricity & 0.09339 & $\emptyset$ \\ 
            Inclination  & 0 & rad\\ 
            Argument of periapsis & $[0,2\pi]$ & rad  \\ 
            Longitude of ascending node & $[0,2\pi]$ & rad \\ 
            True anomaly  & $[0,2\pi]$ & rad \\ \hline
        \end{tabular}
        \tablecomments{The inclination was set to zero, as changing the inclination of the Martian orbit is equivalent to changing its obliquity in a fixed-frame referential centred on Mars. The hooks indicate a range within which a random value is picked.}
        \label{tab:Mars_orb}
    \end{table}
    \begin{figure} [htb!]
      \centering
      \includegraphics[width=0.5
      \textwidth]{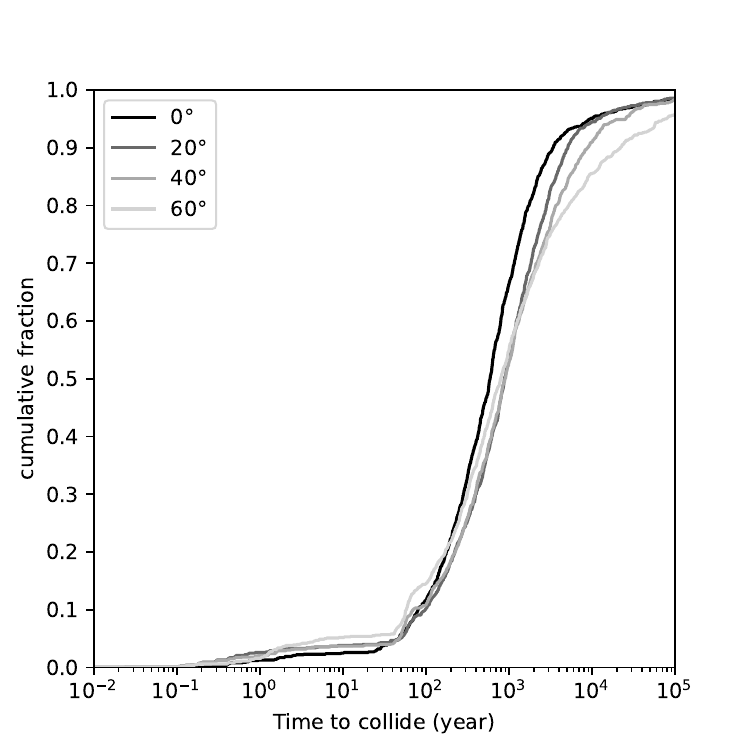}
      \caption{Cumulative collision fraction as a function of time for different values of Mars obliquity. We tried to change Mars obliquity to see if this could affect the probability of collision of post-fragmentation Phobos and Deimos. 1000 simulations were made for each value of the obliquity, and all the physical effects described in this paper were considered. The Martian orbital elements for these simulations are given in Table \ref{tab:Mars_orb}. 
      }
      \label{fig:inclinaisons}
    \end{figure}

    As demonstrated by \cite{Laskar1993}, the obliquity of Mars is chaotic and lies between 0 deg and 60 deg. Therefore, it cannot be asserted that when the potential destruction of the progenitor occurred, the rotation axis of Mars was in a configuration similar to the one of 1950.  A thousand simulations were therefore carried out for different values of Mars obliquity, ranging from 0 to 60 deg, by randomly distributing the semi major axis of Phobos between 5.5 and 6 Mars radii. The orbital parameters of Mars are described in Table \ref{tab:Mars_orb}. As we can no longer use \cite{Ward1979}'s model to compute the revolution of Mars' axis, we now derive it directly from the Sun's position and velocity within the simulations.
        
    Looking at Fig. \ref{fig:inclinaisons}, we can see that a greater obliquity tends to lower the probability of collision. However, even in the most extreme case, the probability of collision remains greater than 94\%. Chaos in Mars' axis of rotation therefore does not alter the results.

    
\subsection{Quantity of fragments} 

    The hypothesis tested by \cite{Hyodo_2022} is the survival of the Phobos and Deimos assuming they are the only two remnants of the fragmentation. However, if a progenitor did fragment, it is conceivable that there were originally more than two fragments, of which Phobos and Deimos would now be the only survivors. It is of interest to verify if the survival probabilities of two objects are altered if the number of fragments is greater than two. To test this hypothesis, we have generated a number $n$ ($n>2$) of objects, of which one has the mass of Phobos, a second has the mass of Deimos, and the others have random masses between those of Phobos and Deimos. The orbital elements of all these objects are then randomly distributed within the ranges given by Table \ref{tab:multi_frag_coll}, which correspond to the maxima and minima of Phobos and Deimos in the results presented by \cite{Bagheri_2021}. To ensure consistency with a fragmentation scenario, all the orbits cross at the ascending node, corresponding to the fragmentation point of the progenitor, and which is located between 5.9 and 6.9 Mars radii. No additional constraints are applied to the orbital element. Note that unlike the two-fragment case, it is not possible to constrain the orbits to have a single point of contact. In order to maximize the fragments' chances of survival, we have considered fusion of fragments after the collision, rather than destructive impacts. The advantage of this assumption is its ability to quickly add up the momentum of the objects, thereby moving them away from each other. 

        \begin{figure} [!t]
          \centering
          \includegraphics[width=0.5
          \textwidth]{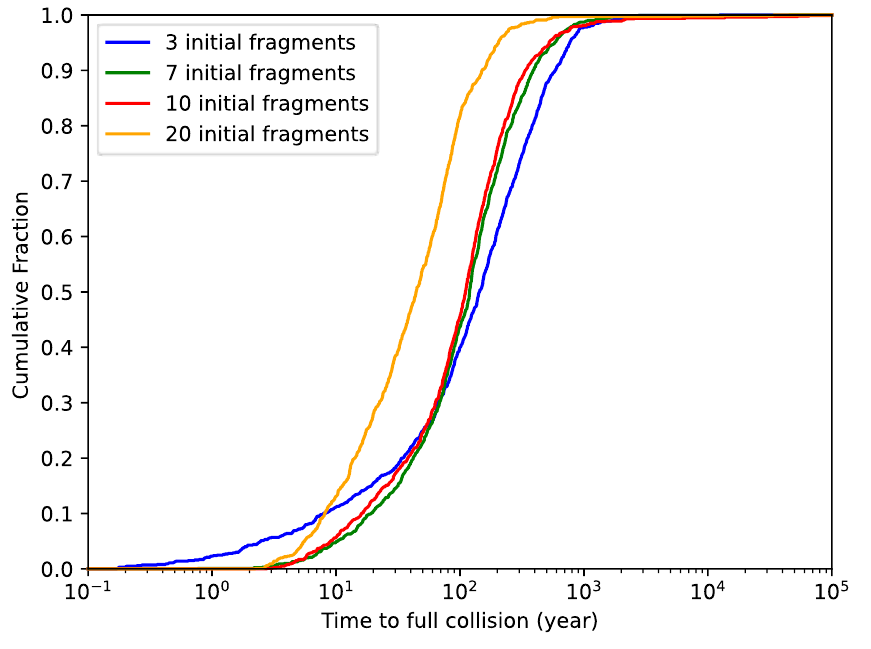}
          \caption{Cumulative fraction of simulations in which all the bodies collided for different number of fragments. We made simulations considering all the physical effects presented in this paper, but we considered that the fragmentation led to more than two fragments. The orbital elements repartitions are given in Table \ref{tab:multi_frag_coll}, and we did 1000 simulations for each number of fragment. We consider that a simulation undergoes "full collision" when all the fragments of the simulations collided with each other. 
          }
          \label{fig:multicoll}
        \end{figure}
    \begin{table}[!htb]
        \centering
        \caption{ Orbital elements of the fragments used in the simulations where more than two fragments were considered. }
        \begin{tabular}{lll}
        \hline \hline
        Orbital element & Value & Unit \\
        \hline
            Semi-major axis  & $[5,7.5]$ & $R_{Mars}$\\ 
            Eccentricity & $[0,0.35]$ & $\emptyset$ \\ 
            Inclination &  $[0.015, 0.021]$ & rad\\ 
            Argument of periapsis & $[0,2\pi]$ * & rad \\ 
            Longitude of ascending node & $[0,2\pi]$ * & rad \\ 
            True anomaly & $[0,2\pi]$ & rad \\ \hline
        \end{tabular}
         \tablecomments{The "*" next to the longitude of ascending node and argument of periapsis values is to notify that unlike other orbital elements, those values are not independent for each of the fragment. Indeed, we assumed that the fragmentation happened at the node, so the longitude of ascending node must be the same for all fragments, and the argument of periapsis must be computed in such a way that all orbits cross each other at the node. This last condition is condition is verified when $\omega = \arccos\left(\frac{a \cdot (1 - \text{e}^2) - d_f}{d_f \cdot e}\right)
    $ with $a$,$e$,$\omega$ the semi-major axis, eccentricity and argument of periapsis of the fragment. $d_f$ is the orbital radius at the fragmentation point, and it is the same for all the fragment.}
        \label{tab:multi_frag_coll}
    \end{table}
    \begin{figure*} [htb!]
      \centering
      \includegraphics[width=1
      \textwidth]{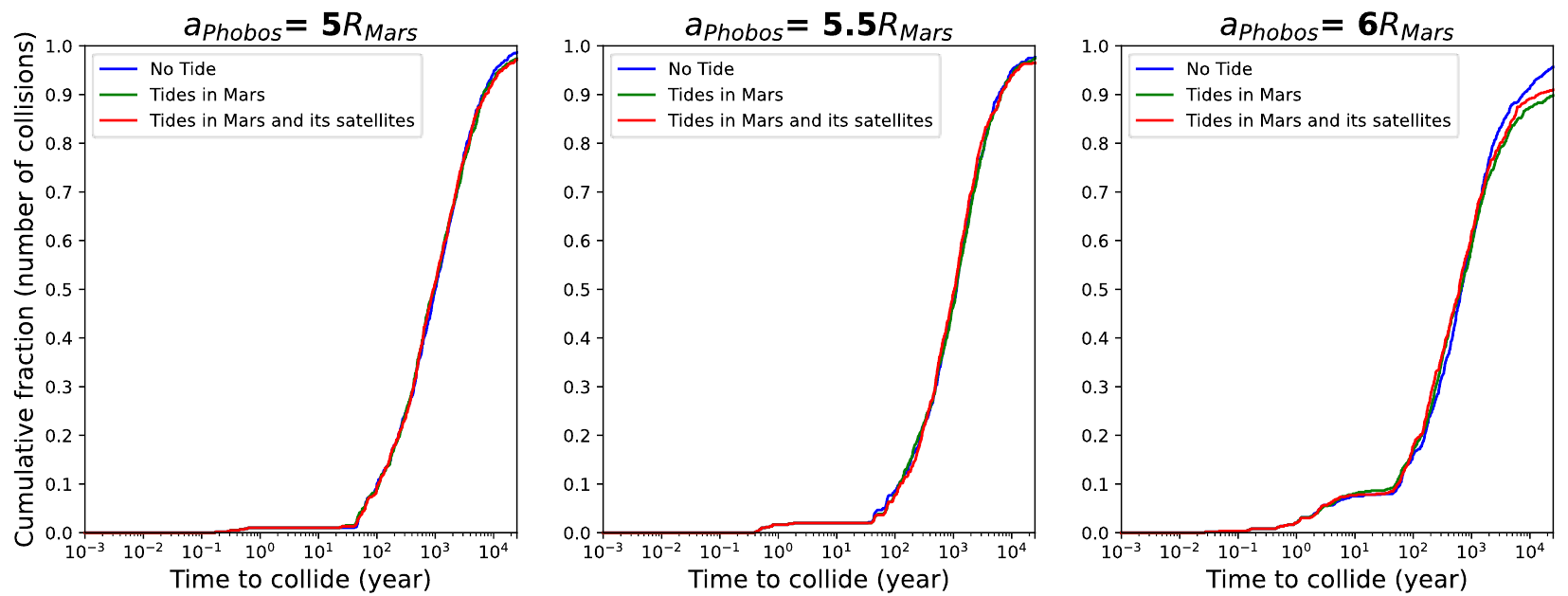}
      \caption{Cumulative fraction of collisions as a function of simulation time, with different values of tides. We compare cumulative fraction of collisions in simulations where we included all the effects presented in this paper. To maximize the effect of the tidal forces, we have considered extreme values of $k_2/Q$ significantly above the current values. It appears that even under such extreme circumstances, the collision rate after 25,000 years remains above 89\% in every case
      } 
      \label{fig:marees}
    \end{figure*}
    The results of these simulations (Fig. \ref{fig:multicoll}) show that there is strictly no case in which at least two fragments survive past 100,000 years.

\subsection{Tidal dissipation in Mars, Phobos and Deimos}\label{sect:tide}

To be exhaustive, we also wanted to verify wether the tidal forces could impact the survival rate of the satellites on such short timescales. As migration due to tidal forces is a very slow phenomenon, we did not expect it to have a strong impact on our results. However, the configuration of Mars and its satellites could have been quite different at the time of their formation (hotter Mars, leading to higher $k_2$ value, different values of $Q$ because of the Phobos and Deimos position after their formation etc.) leading to more important effects of the tidal forces on the satellites orbits than their present values.

To verify if tides could indeed affect the collision rate, we have included them with very high values of $k_2/Q$. We took a value of $0.5$ for Mars $k_2$ and $10$ for its $Q$ while the current values that are used in \cite{Bagheri_2021} are respectively $0.169$ and $95$.
For the same reasons, we also considered values of $k_2/Q$ for Phobos and Deimos that are ten times bigger than those used by \cite{Bagheri_2021}.  All those values are available in Tab. \ref{tab:tide}. 
As we are assuming to be just after a potential fragmentation, we can not assume synchronicity between the moons rotation rate and their revolution angular velocity. For this reason, in each simulation, the initial values of the rotation speed of the moons were randomly chosen between 0 and 10 times Mars' rotation rate. The $c_{22}$ used to compute the change in rotation rate for Phobos and Deimos are respectively 0.016476 (\cite{WILLNER2014}) and 0.04773 (\cite{Huang2024}). 
Integrating the rotation of both Phobos and Deimos was quite demanding in terms of computation time. For this reason, we had to stop the simulation after 25,000 years. 

We can see on Fig. \ref{fig:marees} that the tides in Mars have very few impact on the collision probability. Unexpectedly, the tides inside the bodies seem to have a stronger effect on the collision probability of simulations in which the satellites are further from Mars. This is actually due to a combination of effects : satellites further from Mars in this case are closer to the synchronous orbit, thus are very perturbed by the triaxial shape of Mars. This lowering in the collision probability for these values of semi-major axis is thus due to a combination of the tidal forces and the resonance with Mars rotation. Overall, the extreme tidal forces that we included are barely able to lower the collision probability by a couple of percent after 25,000 years.
     \begin{table}[!t]
        \centering
        \caption{ Tidal values used for the simulations.}
        \begin{tabular}{lll}
        \hline \hline
        Body & $k_2$ & $Q$\\
        \hline
            Mars & 0.5 & 10\\ 
            Phobos & 0.01 & 1000 \\ 
            Deimos  & 0.5 & 1000\\  \hline
        \end{tabular}
        \tablecomments{The values of $k_2$ and $Q$ are not realistic as each of them were chosen to maximize the effect of the tidal forces as explained in Sect.\ref{sect:tide}.}
        \label{tab:tide}
    \end{table}

\section{Conclusion \& Discussions}

As we showed in Sect. \ref{results}, we could not come up with a configuration that lowers the collision probability below 90\%.Scenarios in which Phobos and Deimos could have survived for more than 100,000 years after their formation account for less than 10\% in any of the cases highly conducive to survival that we considered in this study. Thus, the conclusions of \cite{Hyodo_2022} remain reliable, despite some weaknesses in the dynamic modeling of the problem. We want to draw the reader's attention to the fact that these results do not in themselves invalidate the hypothesis proposed by \cite{Bagheri_2021}. As presented in the introduction, there is currently no formation scenario of Phobos and Deimos able to account for all the observations and modeling results of the two moons.
The MMX mission, which will launch in the coming years, will provide further insights into the composition of Mars’ moons (especially Phobos) and will help clarify whether the satellites have an extra-Martian or in-situ origin.

\section*{Data Availability}
The input parameters of each simulation used in this paper are openly available on Zenodo at https://zenodo.org/records/14006551, DOI : \mbox{10.5281/zenodo.14006551}

{
\nolinenumbers
\nolinenumbers
\begin{acknowledgements}
The authors thank Michael Efroimsky for suggesting the idea of testing the results of \cite{Hyodo_2022} by incorporating additional physical effects and his advice throughout this work. The authors also thank Pini Gurfil for his continuous guidance and advice. We would also like to thanks two anonymous reviewers whose comments allowed the authors to greatly improve the manuscript.
\end{acknowledgements}
}


\bibliographystyle{aasjournal}
\bibliography{sources}
\end{document}